\documentclass[aps,prb,twocolumn,showpacs,superscriptaddress]{revtex4}

\usepackage{graphicx}
\usepackage{amsmath}

\begin{document}

\title{Non-superconducting electronic ground state in pressurized BaFe$_2$S$_3$ and BaFe$_2$S$_{2.5}$Se$_{0.5}$}
\author{Hualei Sun}
\affiliation{School of Physics, Sun Yat-Sen University, Guangzhou, Guangdong 510275, China }
\author{Xiaodong Li}
\affiliation{Beijing Synchrotron Radiation Facility, Institute of High Energy Physics, Chinese Academy of Science, Beijing 100049, China}
\author{Yazhou Zhou}
\affiliation{Institute of Physics and Beijing National Laboratory for Condensed Matter Physics, Chinese Academy of Sciences, Beijing 100190, China }
\author{Jia Yu}
\affiliation{School of Physics, Sun Yat-Sen University, Guangzhou, Guangdong 510275, China }
\author{Benjamin A. Frandsen}
\affiliation{Department of Physics and Astronomy, Brigham Young University, Provo, Utah 84602, USA}
\affiliation{Materials Science Division, Lawrence Berkeley National Laboratory, Berkeley, California 94720, USA }
\author{Shan Wu}
\affiliation{Department of Physics, University of California, Berkeley, California 94720, USA }
\affiliation{Materials Science Division, Lawrence Berkeley National Laboratory, Berkeley, California 94720, USA }
\author{Zhijun Xu}
\affiliation{NIST Center for Neutron Research, National Institute of Standards and Technology, Gaithersburg, Maryland 20899, USA}
\affiliation{Department of Materials Science and Engineering, University of Maryland, College Park, Maryland 20742, USA }
\author{Sheng Jiang}
\affiliation{Shanghai Synchrotron Radiation Facility, Zhangjiang Laboratory, Shanghai Advanced Research Institute, Chinese Academy of Sciences, Shanghai 201204, China}
\author{Qingzhen Huang}
\affiliation{NIST Center for Neutron Research, National Institute of Standards and Technology, Gaithersburg, Maryland 20899, USA}
\author{Edith Bourret-Courchesne}
\affiliation{Materials Science Division, Lawrence Berkeley National Laboratory, Berkeley, California 94720, USA }
\author{Liling Sun}
\affiliation{Institute of Physics and Beijing National Laboratory for Condensed Matter Physics, Chinese Academy of Sciences, Beijing 100190, China }
\author{Jeffrey W. Lynn}
\affiliation{NIST Center for Neutron Research, National Institute of Standards and Technology, Gaithersburg, Maryland 20899, USA}
\author{Robert J. Birgeneau}
\affiliation{Department of Physics, University of California, Berkeley, California 94720, USA }
\affiliation{Materials Science Division, Lawrence Berkeley National Laboratory, Berkeley, California 94720, USA }
\author{Meng Wang}
\email{wangmeng5@mail.sysu.edu.cn}
\affiliation{School of Physics, Sun Yat-Sen University, Guangzhou, Guangdong 510275, China }
\begin{abstract}

We report a comprehensive study of the spin ladder compound BaFe$_2$S$_{2.5}$Se$_{0.5}$ using neutron diffraction, inelastic neutron scattering, high pressure synchrotron diffraction, and high pressure transport techniques. We find that BaFe$_2$S$_{2.5}$Se$_{0.5}$ possesses the same $Cmcm$ structure and stripe antiferromagnetic order as does BaFe$_2$S$_3$, but with a reduced N{\'{e}}el temperature of $T_N=98$ K compared to 120 K for the undoped system, and a slightly increased ordered moment of 1.40$\mu_B$ per iron. The low-energy spin excitations in BaFe$_2$S$_{2.5}$Se$_{0.5}$ are likewise similar to those observed in BaFe$_2$S$_{3}$. However, unlike the reports of superconductivity in BaFe$_2$S$_3$ below $T_c \sim 14$~K under pressures of 10~GPa or more, we observe no superconductivity in BaFe$_2$S$_{2.5}$Se$_{0.5}$ at any pressure up to 19.7~GPa. In contrast, the resistivity exhibits an upturn at low temperature under pressure. Furthermore, we show that additional high-quality samples of BaFe$_2$S$_3$ synthesized for this study likewise fail to become superconducting under pressure, instead displaying a similar upturn in resistivity at low temperature. These results demonstrate that microscopic, sample-specific details play an important role in determining the ultimate electronic ground state in this spin ladder system. We suggest that the upturn in resistivity at low temperature in both BaFe$_2$S$_3$ and BaFe$_2$S$_{2.5}$Se$_{0.5}$ may result from Anderson localization induced by S vacancies and random Se substitutions, enhanced by the quasi-one-dimensional ladder structure.

\end{abstract}

\maketitle

\section{Introduction}

The interplay between magnetism and superconductivity has been widely investigated in the copper-oxide and iron-based superconductors, the two known material families comprising the so-called high temperature superconductors ($HTC$s). The mechanism of superconductivity in $HTC$s cannot be explained by the phonon-mediated pairing scenario of conventional BCS theory, motivating the investigation of alternative scenarios in which magnetism plays a crucial role~\cite{Bardeen1957}. Both copper-oxide and iron-based superconductors have layered crystal structures and antiferromagnetically (AF) ordered parent compounds, yet they also have significant differences~\cite{Scalapino2012,Lynn2009,Dai2015,Si2016}. The parent compounds of the copper-oxide superconductors are Mott insulators with split Hubbard bands induced by the strong Coulomb repulsive interaction $U$. On the other hand, the parent compounds of the iron-based superconductors are bad metals with multiple orbitals crossing the Fermi surface. Both the strong and weak correlation scenarios have been employed to describe the iron-based superconductors\cite{Si2016}.

The discovery of superconductivity in BaFe$_2$S$_3$ under pressure has provided a new opportunity for progress in this field\cite{Takahashi2015,Yamauchi2015}. BaFe$_2$S$_3$ is an insulator and exhibits a quasi-one-dimensional (1D) ladder structure (space group $Cmcm$) at ambient pressure~\cite{Chi2016,Wang2016,Zheng2018} that supports stripe-type AF order at low temperature, similar to the 1D copper-oxide ladder system Sr$_{14-x}$Ca$_{x}$Cu$_{24}$O$_{41}$\cite{Uehara1996,Nagata1999,Piskunov2001,Vuletic2006}.  In this sense, BaFe$_2$S$_3$ is a bridge that connects the cooper-oxide and iron-based superconductors, stimulating many theoretical studies of the nature of the insulating state\cite{Rincon2014,Patel2016,Pizarro2019}, the magnetic order\cite{Suzuki2015,Arita2015}, and the insulator-metal transition\cite{Zhang2017,Zhang2018a}.

Interestingly, superconductivity was also discovered in the related compound in BaFe$_2$Se$_3$ under pressure\cite{Ying2017}. BaFe$_2$Se$_3$ crystallizes in the space group $Pnma$ and develops block AF order. The N{\'{e}}el temperature of $T_N\approx240-256$ K and moment size of $M\approx2.8\mu_B$ for  BaFe$_2$Se$_3$ are much larger than that of $T_N\approx104-120$ K and $M\approx1.0-1.3\mu_B$ for BaFe$_2$S$_3$\cite{Saparov2011,Krzton2011a,Lei2011a,Du2012,Caron2012,Nambu2012,Wang2016,Chi2016,Zheng2018}.

In both BaFe$_2$S$_3$ and BaFe$_2$Se$_3$, the effects of electron and hole doping have been extensively explored via substitution of Co or Ni for Fe, and K or Cs for Ba~\cite{Saparov2011,Caron2012,Du2014,Hirata2015,Beom2015,Hawai2017}. It was found that the magnetism can be tuned effectively by doping, but the insulating state is only weakly affected. Other studies have focused on isovalent substitution of Se for S, which induces chemical pressure and acts as a bridge between BaFe$_2$S$_3$ and  BaFe$_2$Se$_3$. Investigations of the BaFe$_2$S$_{3-x}$Se$_x$ system reveal a robust insulating ground state at ambient pressure, but experimental studies of BaFe$_2$S$_{3-x}$Se$_x$ under pressure are still absent~\cite{Du2019,Wu2019}.

In this paper, we report extensive experiments on high-quality samples of BaFe$_2$S$_{2.5}$Se$_{0.5}$ at ambient pressure and under applied pressure. At ambient pressure, BaFe$_2$S$_{2.5}$Se$_{0.5}$ retains the same $Cmcm$ structure as BaFe$_2$S$_3$ and exhibits stripe-type AF order with a reduced N{\'{e}}el temperature of $T_N=98$ K and slightly enhanced moment size of $1.40\pm0.05\mu_B$ compared to BaFe$_2$S$_3$\cite{Takahashi2015,Chi2016,Zheng2018}. The spin gap and spin wave dispersion measured on single crystals of BaFe$_2$S$_{2.5}$Se$_{0.5}$ at ambient pressure are consistent with the Heisenberg model determined from the previous inelastic neutron scattering studies of BaFe$_2$S$_3$ powder samples\cite{Wang2017}. Under applied pressure, we observe an insulator-metal transition around 10~GPa, but superconductivity does not appear at any pressure up to the maximum pressure of $\sim19.7$ GPa reached in this study. Instead, an upturn in resistivity appears at low temperature. We also synthesized high-quality samples of BaFe$_2$S$_3$ in an attempt to reproduce the earlier reports of superconductivity in BaFe$_2$S$_3$ under an applied pressure of $\sim$10~GPa, but superconductivity was not observed in this sample, either. However, BaFe$_2$S$_3$ displays a similar upturn in resistivity at low temperature and high pressure as was seen in BaFe$_2$S$_{2.5}$Se$_{0.5}$. We suggest that this resistivity upturn observed in both materials may be explained by Anderson localization due to the random distribution of S and Se atoms and the presence of vacancies on the (S,Se) sites\cite{Anderson1958}. More generally, the observation that these samples of BaFe$_2$S$_3$ and BaFe$_2$S$_{2.5}$Se$_{0.5}$ show the expected antiferromagnetic insulating ground state at ambient pressure and the pressure-induced insulator-metal transition at $\sim$10~GPa and yet do not display the expected superconducting state demonstrates that microscopic details of specific samples play a decisive role in determining the ground state of this spin ladder system under applied pressure.

\section{Experiment}

High-quality single crystals of BaFe$_2$S$_3$ and BaFe$_2$S$_{2.5}$Se$_{0.5}$ were grown by the Bridgman method. Small Ba chunks, Fe powder, S pieces, and Se shots were loaded in an alumina crucible in an argon-gas-filled glove box with a stoichiometric composition. The alumina crucible was sealed in a quartz tube under vacuum before loading into a box furnace. The sintering process was the same with that of Rb$_{0.8}$Fe$_{1.6}$S$_2$\cite{Wangm2014}. Crystals with a typical size of $2\times 2\times 4$ mm$^3$ were obtained.  Some of the single crystals were ground into powders for subsequent measurements. Neutron powder diffraction (NPD) experiments were carried out on the BT1 powder diffractometer at the NIST Center for Neutron Research (NCNR) using a monochromatic beam with $\lambda = 1.5396$ \AA. A closed cycle refrigerator was used to control the sample temperature. Rietveld refinements of the atomic and magnetic structure were performed using the FullProf Suite\cite{Rodriguez1993}.
Inelastic neutron scattering (INS) measurements were carried out on the BT7 thermal triple axis spectrometer at the NCNR\cite{Lynn2012}.

\emph{In situ}  high-pressure x-ray diffraction (HPXRD) measurements of BaFe$_2$S$_{2.5}$Se$_{0.5}$ were collected at room temperature on the BL15U1 beam line at Shanghai Synchrotron Radiation Facility using a diamond anvil cell and x-rays with an energy of 20 keV ($\lambda=0.6199$ \AA). This energy is sufficiently high to pass through diamond anvil cells. A two dimensional detector was used to record the diffraction pattern with an exposure time of 60~s. A pair of symmetric diamond anvils with a 300 $\mu$m culet was used to apply the pressure. A steel gasket surrounded the sample chamber of diameter 120 $\mu$m. A pre-compressed pellet was loaded in the middle of the sample chamber with silicone oil acting as the pressure transmitting medium. The pressure in the diamond anvil cell was determined by measuring the shift of the fluorescence wavelength of the ruby spheres that were placed inside the sample chamber\cite{Mao1986}. The data were initially processed using Fit2d with a CeO$_2$ calibration, and subsequent Le Bail refinements were performed using the GSAS software\cite{Larson1994}.

\emph{In situ}  high-pressure electrical resistance measurements on BaFe$_2$S$_3$ and BaFe$_2$S$_{2.5}$Se$_{0.5}$ single crystals were carried out in diamond anvil cells made from a Be-Cu alloy using a standard four-probe technique. The diamond anvil cell for BaFe$_2$S$_{2.5}$Se$_{0.5}$ was loaded into a house-built refrigerator for temperature control. The BaFe$_2$S$_3$ sample was measured using a physical property measurement system (PPMS) (Quantum Design). Diamond anvils with a 400 $\mu$m culet and a steel gasket were used for the sample chamber of diameter 150 $\mu$m. Insulation from the gasket was achieved with a thin layer of a mixture of cubic boron nitride and epoxy. NaCl powders were employed as the pressure transmitting medium. The pressures in the resistance measurements were calibrated by the ruby fluorescence shift at room temperature.

\section{Results}
\subsection{Neutron diffraction}
\begin{figure}[b]
\includegraphics[scale=0.3]{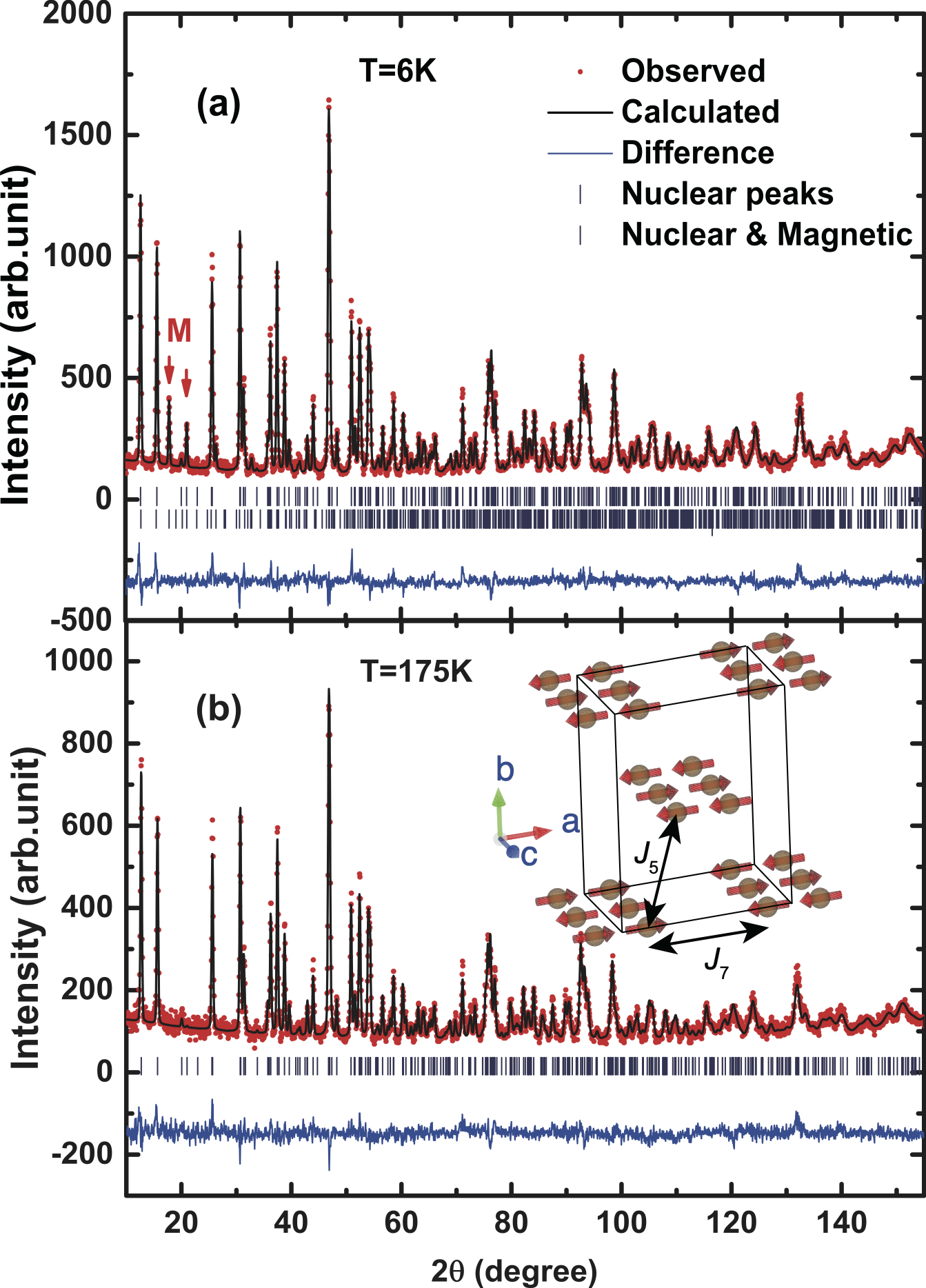}
\caption{ Neutron diffraction spectra of BaFe$_2$S$_{2.5}$Se$_{0.5}$ collected at (a) 6 and (b) 175 K. The black solid curves are calculated from the refined structures, and the blue curves show the fit residuals. The inset in (b) is a sketch of the structure with representative magnetic exchange interactions $J_5$ and $J_7$.}
\label{fig1}
\end{figure}

\begin{table}[t]
\caption{Structural parameters of BaFe$_2$S$_{2.5}$Se$_{0.5}$ at 6 K. The space group is $Cmcm$ (No. 63) and the refined lattice constants are $a=8.8070(2), b=11.2624(2), c=5.2868(1)$ \AA. The agreement factors are $R_p=15.31\%, \omega R_p=16.39\%, \chi^2=1.30\%$.}
\begin{tabular}{p{1cm}p{1cm}p{1.5cm}p{1.5cm}p{1cm}p{1.5cm}}
\hline \hline
Atom & Site & x         & y         & z    & Occ.     \\ \hline
Ba   & 4c   & 0.5       & 0.1879(4) & 0.25 & 1        \\
Fe   & 8e   & 0.3475(2) & 0.5       & 0    & 0.99(1)        \\
S1   & 4c   & 0.5       & 0.6158(7) & 0.25 & 0.811(5) \\
Se1  & 4c   & 0.5       & 0.6158(7) & 0.25 & 0.131(5) \\
S2   & 8g   & 0.2031(4) & 0.3777(4) & 0.25 & 0.81(2) \\
Se2  & 8g   & 0.2031(4) & 0.3777(4) & 0.25 & 0.19(2) \\ \hline \hline
\end{tabular}
\label{table1}
\end{table}

\begin{figure}[b]
\includegraphics[scale=0.3]{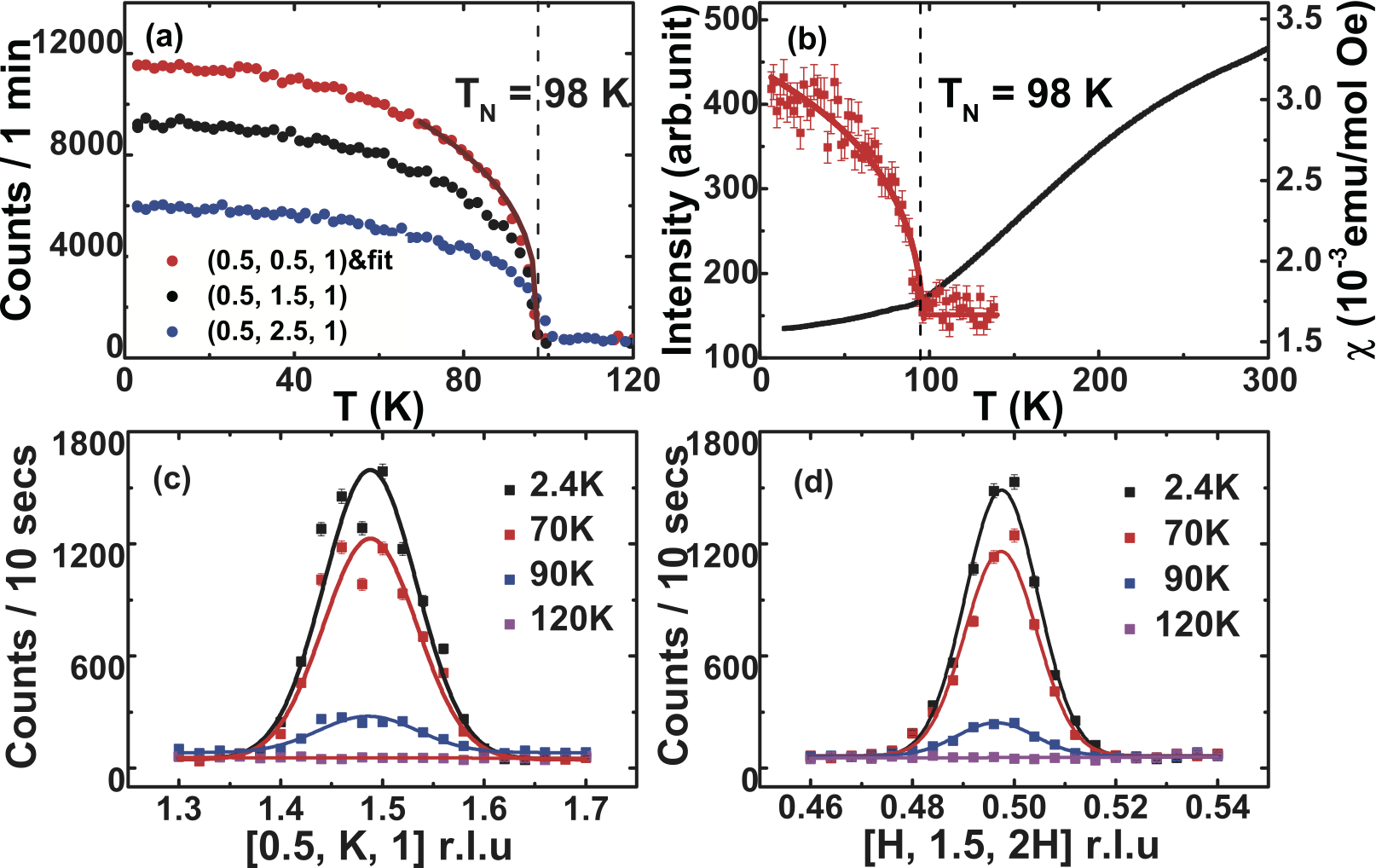}
\caption{ (a) Intensities of the magnetic peaks at $Q=$(0.5, 0.5, 1), (0.5, 1.5, 1), and (0.5, 2.5, 1) as a function of temperature for a BaFe$_2$S$_{2.5}$Se$_{0.5}$ single crystal. (b) Magnetic order parameter of a powder sample measured at $2\theta=24.23^\circ$, which corresponds to the wave vector $Q=(0.5, 0.5, 1)$. The black solid curve is a dc magnetic susceptibility measured with 1 T field along the ladder direction. The purple curve in (a) is a fit using a power-law form $\phi(T)^2\propto(1-T/T_N)^{2\beta}$, resulting in $\beta=0.17\pm0.01$.  The red line in (b) is a guide to the eyes. (c) Magnetic peak scans along the $[0.5, K, 1]$ direction at 2.4, 70, 90, and 120 K. The magnetic peak disappears at 120 K. (d) Identical scans along the $[H, 1.5, 2H]$ direction. The solid lines in (c) and (d) are Gaussian fits to the experimental data. The error bars represent one standard deviation of the measured counts throughout this paper. Note: 1 emu/(mol Oe) = 4$\pi\times10^{-6}$ m$^3$/mol. }
\label{fig2}
\end{figure}

NPD patterns collected from a powder sample of BaFe$_2$S$_{2.5}$Se$_{0.5}$ at 6 and 175 K at ambient pressure are plotted in Fig. \ref{fig1}. The nuclear Bragg peaks are well described by the $Cmcm$ structure exhibited by BaFe$_2$S$_3$. The refined parameters of the nuclear structure are listed in Table \ref{table1}. We found that vacancies occur randomly throughout the crystal on 6$\%$ of the (S,Se) sites (the $4c$ Wyckoff position). Additional peaks present at 6 K but not at 175 K (marked by the red arrows and letter M) indicate the presence of long-range antiferromagnetic order, which is well fit by the stripe-type pattern found in BaFe$_2$S$_3$. However, the ordered moment size increases from $1.29\pm0.03$ in the pure S compound to $1.40\pm0.05\mu_B$ in the present compound\cite{Zheng2018}.

To investigate the N{\'{e}}el temperature of BaFe$_2$S$_{2.5}$Se$_{0.5}$, we measured the magnetic peak intensities at $Q=(H, K, L)=(0.5, 0.5, 1), (0.5, 1.5, 1)$, and $(0.5, 2.5, 1)$ as a function of temperature for single crystal samples, as shown in Fig. \ref{fig2}(a). Here, $(H, K, L)$ are Miller indices for the momentum transfer $|Q|=2\pi\sqrt{(H/a)^2+(K/b)^2+(L/c)^2}$, where the lattice constants are $a=8.8070(2), b=11.2624(2)$, and $c=5.2868(1)$ \AA. The temperature dependence of the magnetic peak intensities indicates a N{\'{e}}el temperature of $T_N=98\pm2$ K, which is lower than $T_N=120$~K for BaFe$_2$S$_{3}$~\cite{Zheng2018}. A simple power-law dependence $\phi(T)^2\propto(1-T/T_N)^{2\beta}$ is employed to fit the peak intensity at $Q=(0.5, 0.5, 1)$ between $0.7\leq T/T_N\leq 1$, resulting in $\beta=0.17\pm0.01$. If we fit the data between $0.9\leq T/T_N\leq 1$, the result is $\beta=0.21\pm0.01$. This value is close to that of LaFeAsO, which is between $\beta=0.125$ of the 2D Ising model and $\beta=0.326$ of the 3D Ising model\cite{Guillou1980,Wilson2010}. In Fig.~\ref{fig2}(b), we overlay the NPD intensity of the $(H, K, L)=(0.5, 0.5, 1)$ magnetic Bragg peak with the magnetic susceptibility, both of which reveal an identical N{\'{e}}el temperature of $98$~K. Additionally, we aligned a single crystal in the $[H, K, 2H]$ scattering plane and scanned both along the $[0.5, K, 1]$ [Fig. \ref{fig2}(c)] and $[H, 1.5, 2H]$ directions [Fig. \ref{fig2}(d)] at $T=2.4, 70, 90$, and 120 K. The magnetic peaks disappear above $T_N$.  The magnetic peaks have a full width at half maximum (FWHM) close to the instrumental resolution, demonstrating that the magnetic structure is long-range ordered in BaFe$_2$S$_{2.5}$Se$_{0.5}$.

\subsection{Spin excitations}
\begin{figure}[t]
\includegraphics[scale=0.65]{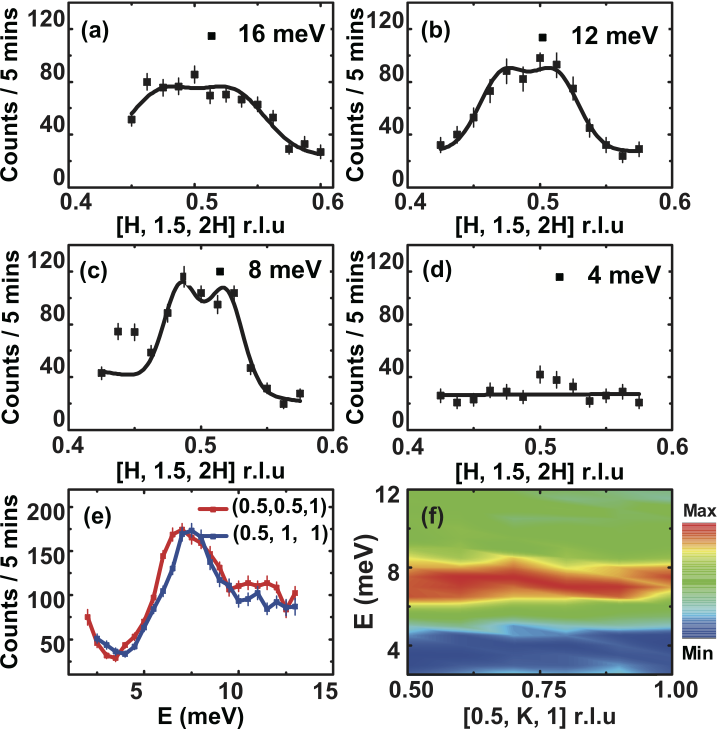}
\caption{ Constant energy scans along the $[H, 1.5, 2H]$ r.l.u direction for (a) $\Delta E$=16, (b) 12, (c) 8, and (d) 4 meV. The solid curves in (a-c) are fits using two Gaussian peaks. The line in (d) is a linear fit. (e) Constant $Q$ scans at (0.5, 0.5, 1) and (0.5, 1, 1). (f) A color map of the spin excitations covering half the Brillouin zone along the $[0.5, K, 1]$ direction. All data were collected at 5 K.}
\label{fig3}
\end{figure}

Magnetic exchange interactions are widely believed to be intimately related to the mechanism of $HTC$s. The spin waves of the stripe AF order have previously been investigated in powder samples of BaFe$_2$S$_3$, revealing strong intra-ladder ferromagnetic exchange interactions along the rung direction $SJ_R=-71\pm4$ meV, antiferromagnetic couplings along the leg direction $SJ_L=49\pm3$ meV, and inter-ladder couplings along the $a$ direction $SJ_7=3.0\pm0.5$ meV\cite{Wang2017}, as illustrated in the inset of Fig. \ref{fig1}(b). However, weak inter-ladder couplings along the $b$ direction, $SJ_5$, could not be deduced. The spin excitations of BaFe$_2$S$_{2.5}$Se$_{0.5}$ were measured on a single crystal of mass 0.25 g aligned in the $[H, K, 2H]$ scattering plane. The results are presented in Fig. \ref{fig3}. A spin gap of 5 meV and clear dispersions are observed along the $[H, 1.5, 2H]$ direction as shown in Figs. \ref{fig3}(a)$-$\ref{fig3}(e). The dispersions are consistent with the Heisenberg Hamiltonian deduced from BaFe$_2$S$_3$. To check the inter-ladder magnetic exchange interactions along the $b$ direction, $SJ_5$, we show the constant $Q$ scans from the Brillioun zone center at $Q=(0.5, 0.5, 1)$ to the Brillioun zone corner at $Q=(0.5, 1, 1)$ in Figs. \ref{fig3}(e) and \ref{fig3}(f). No spin wave dispersion is observed within the resolution of our instrument. An estimation of the exchange interaction ($SJ_5$) between the ladders along the $b$ direction would be smaller than 1.5 meV.

\subsection{High pressure studies of BaFe$_2$S$_{2.5}$Se$_{0.5}$}

\begin{figure}[t]
\includegraphics[scale=0.42]{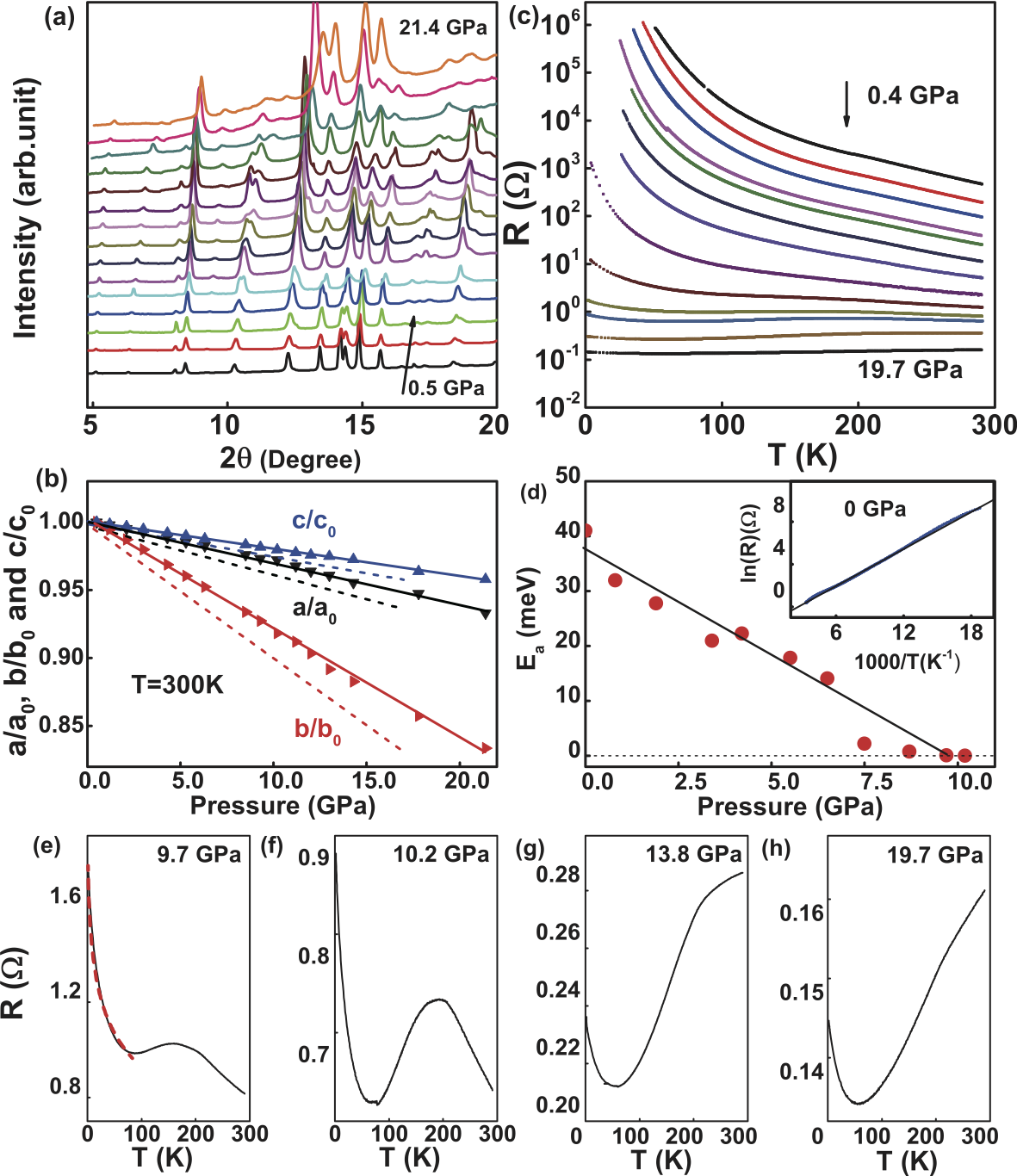}
\caption{(a) HPXRD measurements on a powder sample of BaFe$_2$S$_{2.5}$Se$_{0.5}$ under pressure. (b) Pressure dependence of the lattice constants for BaFe$_2$S$_{2.5}$Se$_{0.5}$ (solid lines with data points) and BaFe$_2$S$_3$ from literature normalized by their values at ambient pressure\cite{Takahashi2015}. (c) Resistance of BaFe$_2$S$_{2.5}$Se$_{0.5}$ as a function of temperature for different pressures from 0.4 to 19.7 GPa, displayed on a logarithmic vertical scale. (d) The fitted thermal activation gap as a function of pressure. The solid line is a guide to the eye. Inset: Scaled resistance and best fit at ambient pressure. (e-h) Temperature dependence of the resistance under applied pressures of (e) 9.7, (f) 10.2, (g) 13.8, and (h) 19.7 GPa. No superconductivity is observed. The red dashed line is a fit using $R=R_0+A\times$ln$(1/T)$, where $A$ is a prefactor.}
\label{fig4}
\end{figure}

To study the properties of BaFe$_2$S$_{2.5}$Se$_{0.5}$ under pressure, we measured the structure at 300 K using x-ray diffraction with \emph{in situ} pressure up to 21.4~GPa, as well as the resistance between 2~K and 300~K at various pressures up to 19.7~GPa. From the HPXRD scans shown in Fig.~\ref{fig4}(a), no structural transition is observed up to 21.4 GPa. The diffraction spectra are affected by the orientation of the sample and broadening of the peaks under pressure. In Fig.~\ref{fig4}(b), we plot the ratios of the compressed lattice constants to the lattice constants at ambient pressure. These ratios decrease with pressure more slowly in BaFe$_2$S$_{2.5}$Se$_{0.5}$ than in BaFe$_2$S$_3$, consistent with the expectation for Se substitution. Upon increasing the pressure from 0.4 to 19.7 GPa, the electrical resistance over the measured temperature range decreases by 7 orders of magnitude as seen in Fig.~\ref{fig4}(c), representing a transition from an insulator to a metal. The thermal activation gap $E_a$ obtained from fitting the resistance curves using the empirical function $R=R_0exp(E_a/k_BT)$ decreases monotonically and closes around 10 GPa [Fig. \ref{fig4}(d)]. However, in contrast to the subsequent appearance of superconductivity in BaFe$_2$S$_3$ reported in the literature, we did not observe superconductivity in BaFe$_2$S$_{2.5}$Se$_{0.5}$ up to 19.7 GPa~\cite{Takahashi2015,Yamauchi2015}. The resistance curves of BaFe$_2$S$_{2.5}$Se$_{0.5}$ at 9.7, 10.2, 13.8, and 19.7 GPa are presented in Figs. \ref{fig4}(e)$-$\ref{fig4}(h). An insulator-metal transition could be identified as the existence of the humps in the resistance around 180 and 200 K for 9.7 and 10.2 GPa, respectively, as seen in Figs. \ref{fig4}(e) and \ref{fig4}(f). The hump in resistance moves outside of the measurable temperature range for the higher pressures. The resistance decreases with decreasing temperature down to $\sim$50 K, yielding a metallic feature in Figs. \ref{fig4}(g) and \ref{fig4}(h). However, an obvious upturn in resistance emerges at low temperatures for all pressures shown in panels (e-h) of Fig.~\ref{fig4}. This upturn should not be attributed to a proper insulating state, considering that the increase of resistance down to the lowest temperature of 2 K is unlike the empirical behavior $R=R_0exp(E_a/k_BT)$ expected for an insulator or a semiconductor. Instead, the resistance at low temperatures follows a logarithmic function.

\subsection{High pressure studies of BaFe$_2$S$_3$}

\begin{figure}[t]
\includegraphics[scale=0.5]{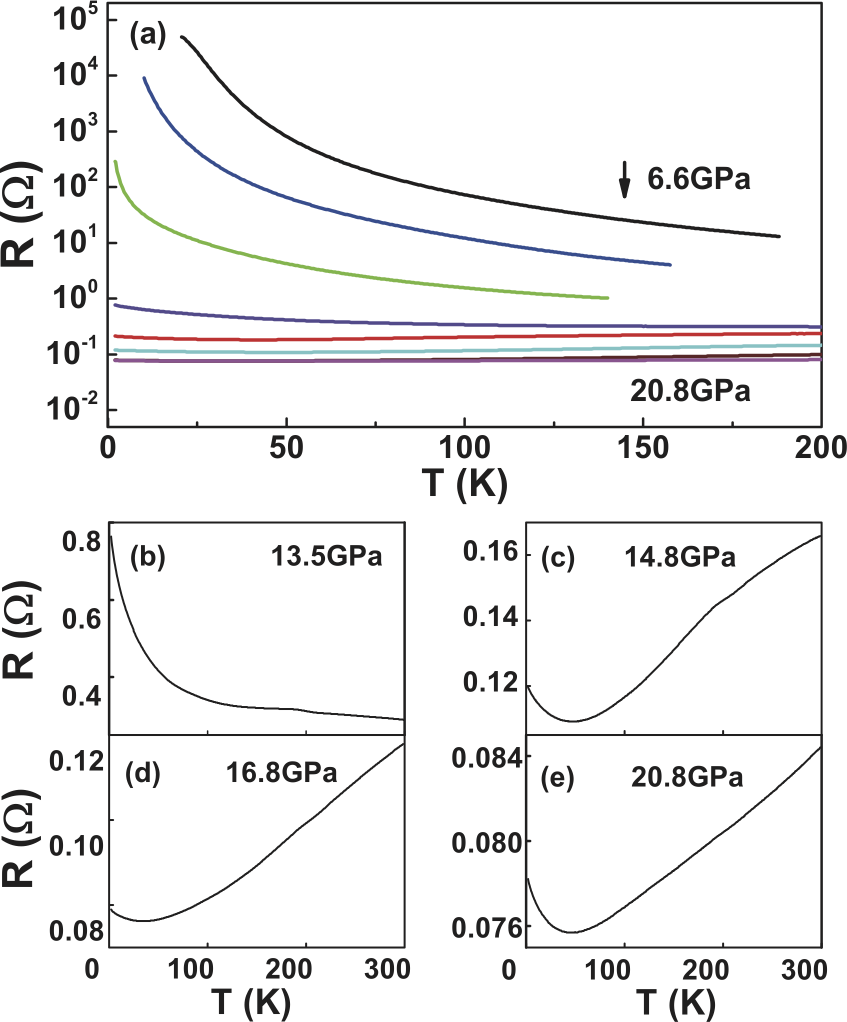}
\caption{ Resistance of BaFe$_2$S$_3$ as a function of temperature for pressures from 6.6 to 20.8 GPa on a logarithmic scale.  (b-e) Representative resistance of BaFe$_2$S$_3$ at 13.5, 14.8, 16.8, and 20.8 GPa on a linear scale.}
\label{fig5}
\end{figure}

The structural and magnetic properties of BaFe$_2$S$_{2.5}$Se$_{0.5}$ are similar to those of BaFe$_2$S$_3$. To understand why superconductivity does not appear in our sample of BaFe$_2$S$_{2.5}$Se$_{0.5}$ under pressure, we also measured the pressure-dependent resistance of a sample of BaFe$_2$S$_3$ synthesized using the same procedure~\cite{Wang2017,Zheng2018}. The BaFe$_2$S$_3$ sample we used has a slightly larger moment size of $1.29\pm0.03\mu_B$ and a more energetically stable position of the S atoms at the 8$g$ Wyckoff sites compared to published works reporting superconductivity in BaFe$_2$S$_3$\cite{Takahashi2015,Chi2016}. The resistance as a function of temperature under different pressures is presented in Fig. \ref{fig5}. An insulator-metal transition occurs around 10~GPa, consistent with previous studies on BaFe$_2$S$_3$ and our measurements on BaFe$_2$S$_{2.5}$Se$_{0.5}$\cite{Takahashi2015,Yamauchi2015}. The resistance curves for representative pressures of 13.5, 14.8, 16.8, and 20.8 GPa are plotted in Figs. \ref{fig5}(b)$-$\ref{fig5}(e). No superconductivity is observed, differing from the previous reports for BaFe$_2$S$_3$. The similarity of the resistance curves of BaFe$_2$S$_3$ and BaFe$_2$S$_{2.5}$Se$_{0.5}$ under pressure indicates that the mechanism of the upturn, the insulator-metal transition, and the absence of superconductivity in the two compounds likely have the same origin.

 \section{Discussion and Summary}

The samples of BaFe$_2$S$_{3}$ and BaFe$_2$S$_{2.5}$Se$_{0.5}$ used in this study have the expected $Cmcm$ structure and stripe antiferromganetic order at ambient pressure and undergo the expected pressure-driven insulator-metal transition around 10~GPa, yet neither sample has a superconducting ground state at high pressure. This raises the question of why some samples of BaFe$_2$S$_3$ become superconductors at high pressure while others do not. Previous studies have reported sample-dependent variations in the shape of the resistance curve, magnetic ordering temperature, ordered magnetic moment, and unit cell dimensions at ambient pressure, likely as a result of small differences in stoichiometry based on the synthesis procedure~\cite{Yamauchi2015,Hirata2015,Zhang2018,Zheng2018}. Such effects have also been seen in the related compound BaFe$_2$Se$_3$~\cite{Krzton2011a,Hawai2017,Ying2017}. The present work indicates that sample-dependent properties also carry over to the pressure-induced superconducting state, which appears to be quite delicate and sensitive to microscopic details that may otherwise be overlooked. On the other hand, the pressure-driven insulator-metal transition seems to be a more robust feature of these spin ladder systems, as it is observed in samples both with and without superconductivity at high pressure. This transition may be attributable to the increase of $W/U$, where $W$ is the electronic bandwidth and $U$ is the Coulomb repulsion, or to the enhancement of the quasiparticle weight near the Fermi surface\cite{Materne2019,Pizarro2019}.

In BaFe$_2$S$_{2.5}$Se$_{0.5}$, the obtained magnetic critical exponent of $\beta=0.21\pm 0.01$ is in-between the 2D and 3D Ising models, indicating that BaFe$_2$S$_{2.5}$Se$_{0.5}$ is a compound in the crossover region. This is supported by the inelastic neutron scattering results which reveal a coupling that could not be detected in our measurements ($SJ_5<1.5$ meV) along the $b$ direction. The magnetic couplings could be decreased due to the expansion of  the lattice constants via Se doping. A 3D Ising transition is necessary to be investigated in nearby compounds of BaFe$_2$S$_{2.5}$Se$_{0.5}$.

As has been established, the high-pressure ground state of our samples of BaFe$_2$S$_{3}$ and BaFe$_2$S$_{2.5}$Se$_{0.5}$ is not superconducting, yet it is not a conventional metal either. The upturn in resistance observed in both samples at low temperature and high pressure is a clear deviation from typical metallic behavior, but it is also significantly different from the exponential activated behavior expected for a typical insulator or a semiconductor. Instead, the resistance curve follows a logarithmic function, which is consistent with Anderson localization due to disorder in the system~\cite{Ying2016}. Indeed, such a scenario of Anderson localization has been proposed for hole-doped Ba$_{1-x}$Cs$_x$Fe$_2$Se$_3$\cite{Hawai2017}. In the present case, disorder due to random S/Se mixing, 6\% vacancies on the S/Se $4c$ Wyckoff sites, and a distribution of positions of the S/Se atoms on the $8g$ Wyckoff sites are potential causes of the Anderson localization\cite{Chi2016,Zheng2018}. This scenario can explain why the resistance upturn is larger in BaFe$_2$S$_{2.5}$Se$_{0.5}$ than in BaFe$_2$S$_3$ [compare Figs.~\ref{fig4}(e)$-$\ref{fig4}(h) and \ref{fig5}(b)$-$\ref{fig5}(e)], since the S and Se mixing provides additional disorder. Furthermore, the low dimensionality of this system would enhance the Anderson localization effect\cite{Alloul2009}. We suggest that Anderson localization may therefore be a competing tendency in this spin ladder system that can suppress superconductivity in samples with an increased level of disorder.

In summary, we have measured the crystal structure, magnetic order, and low-energy spin excitations of BaFe$_2$S$_{2.5}$Se$_{0.5}$ at ambient pressure, as well as the pressure dependence of the structure and resistance up to high pressures of $\sim$20~GPa. The N{\'{e}}el temperature of $T_N=98$ K, ordered moment size of $1.40\pm0.05\mu_B$/Fe, and thermal activation gap have been determined. The inter-ladder magnetic exchange interaction $SJ_5$ along the $b$ axis is estimated to be smaller than 1.5 meV. Superconductivity does not appear at high pressure in this sample of BaFe$_2$S$_{2.5}$Se$_{0.5}$, but instead, an upturn in resistance potentially attributable to Anderson localization has been observed in the low-temperature, high-pressure state. A pressure study of the resistance of pure BaFe$_2$S$_3$ also reveals the absence of superconductivity and the presence of this potential Anderson localized state in the high-pressure regime, in contrast to previous reports of superconductivity in BaFe$_2$S$_3$. These results demonstrate that the appearance of superconductivity in this 1D ladder system not only depends on the electronic correlation and spin fluctuations, but is also sensitive to sample-specific details such as precise stoichiometry and microstructure that may often be overlooked, with Anderson localization as a possible pathway to an alternate, non-superconducting ground state. Careful attention should be given to the sample dependence of the superconducting properties in future studies of this system.

 \section{Acknowledgement}

Work at Sun Yat-Sen University was supported by the National Natural Science Foundation of China (Grant No. 11904414, 11904416), Natural Science Foundation of Guangdong  (No. 2018A030313055), National Key Research and Development Program of China (No. 2019YFA0705700), the Fundamental Research Funds for the Central Universities (No. 18lgpy73),  the Hundreds of Talents program of Sun Yat-Sen University, and Young Zhujiang Scholar program. Work at Institute of Physics of CAS was supported by the National Key Research and Development Program of China (No. 2017YFA0302900, 2016YFA0300300). Work at UC Berkeley and Lawrence Berkeley Laboratory was supported by the Office of Science, Office of Basic Energy Sciences (BES), Materials Sciences and Engineering Division of the U.S. Department of Energy (DOE) under Contract No. DE-AC02-05-CH1231 within the Quantum Materials Program (KC2202) and BES. The identification of any commercial product or trade name does not imply endorsement or recommendation by the National Institute of Standards and Technology.


%\bibliography{NoEndingPoint}

\bibliography{mengbib}

\end{document}